\documentclass[twocolumn]{aastex6}
\usepackage{amsmath,amsfonts,amsthm} 
\usepackage{graphicx} 
\usepackage{bmpsize}
\usepackage{color,soul}

\fullcollaborationName{The OATo Relativistic Astrometry Collaboration}

\begin{document}


\title{The Short Term Stability of a Simulated Differential Astrometric Reference Frame in the Gaia era}

\author{Ummi Abbas\altaffilmark{1}, Beatrice Bucciarelli, Mario G. Lattanzi, 
Mariateresa Crosta, Mario Gai, Richard Smart, Alessandro Sozzetti, Alberto Vecchiato}
\affil{Osservatorio Astrofisico di Torino \\ 
Via Osservatorio 20 \\ 
Pino Torinese, I-10025, Italy}

\altaffiltext{1}{abbas@oato.inaf.it}

\begin{abstract}

We use methods of differential astrometry to construct a small field 
inertial reference frame stable at the micro-arcsecond level.
Such a high level of astrometric precision can be expected with the end-of-mission standard errors
to be achieved with the Gaia space satellite using global astrometry. 
We harness Gaia measurements of field angles 
and look at the influence of the number of reference stars and the star's magnitude as well as 
astrometric systematics 
on the total error budget 
with the help of Gaia-like simulations around the Ecliptic Pole in a differential astrometric scenario. 

We find that the systematic errors are modeled and reliably estimated to the $\mu$as level even in 
fields with a modest number of 37 stars with G $<$13 mag over a $0.24$ sq.degs. field of view
for short time scales of the order of a day with high-cadence observations such as those 
around the North Ecliptic Pole during the EPSL scanning mode of Gaia for a perfect 
instrument. The inclusion of the geometric instrument model over such short time scales accounting
for large-scale calibrations requires fainter stars down to G = 14 mag without 
diminishing the accuracy of the reference frame. 
We discuss several future perspectives of utilizing this methodology
over different and longer timescales. \footnote{This is an author-created, un-copyedited version of an 
article accepted for publication in Publications of the Astronomical Society of the Pacific. The publisher 
is not responsible for any errors or omissions in this version of the manuscript or any version derived from it. 
The Version of Record is available online at [DOI TBC].}

\end{abstract}

\keywords{astrometry -- methods: data analysis -- methods: statistical -- reference systems}

\section{Introduction} \label{sec:intro}

The principles of differential astrometry can be used in a variety of applications: e.g. 
for improving the origin of a fundamental star-catalog coordinate system 
\citep{zverev1976} with the help of differential astrometric observations of the Sun and planets;
cluster membership studies that can be performed using relative proper motions and even 
in deriving absolute proper motions of globular clusters and distant open clusters \citep{vanAltena2013,
dinescu1997};
detecting the reflex motion of the target star due to the presence of its planets 
\citep{sozzetti2005}; obtaining trigonometric parallaxes 
\citep{benedict2002a, benedict2009, mcArthur2011, mcArthur1999, riess2014, casertano2016} 
or measuring the relativistic deflection due 
to the various moments of massive planets and in determining PPN-$\gamma$ \citep{crosta2006}. 

The preliminary step before studying such `differential' effects involves the establishment 
of a reliable `local' inertial reference frame which we attempt in this paper.
In general, a `global' reference frame is defined by the positions of objects through their coordinates 
in a reference system (a coordinate system specifying the direction of the axes and the zero point 
or the origin) and thereby represents a practical realization of the reference system 
\citep{johnston1999}. 
For example, the International Celestial Reference Frame (ICRF) is the realization of the International Celestial Reference System (ICRS) and thereby defines the directions of its axes.
At different wavelengths the precision of the ICRS-axes orientation varies reaching $\sim$10 
micro-arcsecond ($\mu$as) at radio frequencies, leading to the second realization of the ICRF, i.e. 
ICRF2, and milli-arcseconds (mas) at optical wavelengths. 
At radio frequencies, the ICRF is defined mainly by very distant extragalactic sources having no discernible proper motions leading to the definition of a quasi-inertial reference frame.
On the other hand, for small fields ($\sim$ 1 square degree on the sky) the local reference frame is 
inertial in the sense that the positions and motions of an object that are due to forces acting 
on the objects can be reliably modeled within the random measurement errors \citep{treuhaft1990}.

It is desirable to use many fixed and stable points to define a reliable local reference 
frame even though in principle a couple of points are enough, e.g. right ascension and declination describing the 
angular distance from the origin of right ascension and from the catalog equator respectively \citep{johnston1999}.
A modern astrometric catalog contains data on a large number of objects, 
so the coordinate system is vastly overdetermined but allows to decrease the `random' measurement error by 
a factor of the square root of the total number of observations.
The rms scatter of repeated observations of the positions 
of these objects in a given reference frame define its stability.  
Typically, every catalog contains systematic errors, i.e. errors in position that are similar in 
direction and magnitude for objects sharing similar characteristics, lying in the same area of the sky, 
or are of the same magnitude (flux) or color (spectral index). 
Systematic errors lead to biases in the reference frame that is effectively different for 
different classes of objects. Obviously, minimizing systematic errors through good physical models when a 
catalog is constructed is as important, if not more so, than minimizing the random errors. 

In the presence of Earth's atmosphere, limitations to the astrometric precision are caused by effects 
such as refraction, turbulence, delays, etc \citep{sozzetti2005}. In its absence, for space-based measurements and for those 
that are differential in nature (based on reference objects that are all within a small field), 
we need to address effects such as: light aberration that is of the order of $\sim$20 
arcseconds to first order and a few mas to second order \citep{klioner2003};  
gravitational deflection terms that lead to effects of several mas even at the Ecliptic Pole due to 
the monopole moment of the Sun \citep{crosta2006, turyshev2002}; parallaxes and proper 
motions of stars that can be either removed apriori or accounted for in the model \citep{vanAltena2013};
and for changes in the geometric instrument model due to thermal variations and imperfections in the 
instrument that need to be efficiently calibrated \citep{lindegren2012}.

In this paper we use Gaia-like simulations that are optimized for global astrometry and study the
measurements with a differential astrometric approach. We analyze the efficiency of such a 
method in constructing a high-precision inertial astrometric reference frame over small fields 
whose size is determined by the dimensions of Gaia's field of view. We describe the total error budget 
in terms of various systematics due to different astrometrical effects, astrophysical effects due 
to the star's magnitude, and geometrical effects due to the distribution of a given number of stars.

The paper is divided into the following sections: In Sec.~\ref{sec:simulation} we describe the details 
of the simulation set-up, Sec.~\ref{sec:limitations} presents the properties of a space-based 
differential reference frame, the following Sec.~\ref{sec:principles} discusses the principles of Differential Astrometry 
alongwith the applied astrometric modeling in Sec.~\ref{sec:models} and some 
of the results obtained (Sec.~\ref{sec:residuals}) and their Discussion (Sec.~\ref{sec:disc})
finally wrapping up with Future Perspectives (Sec.~\ref{sec:applic}).

\section{The simulation set-up} \label{sec:simulation}

The Gaia space satellite will perform unprecedented high precision global astrometry 
at the $\mu$arcsecond level of roughly 1 billion stars down to G $\sim$ 20 mag.
Under such conditions we want to study the potential of differential 
astrometric measurements in establishing a local reference frame. 
We use simulated Gaia observations as a testbed for our analysis, which however is general enough to be 
applied to any scenario with overlapping measurements that can be treated in a differential
manner ultimately leading to a reliable local Reference Frame. 
We attempt this by using the field angles, $\eta$ and $\zeta$, as measured in 
the field of view (FOV) of Gaia respectively in the scanning direction of the satellite and 
perpendicular to it 
instead of using global coordinates, i.e. positions ($\alpha$, $\delta$),  
that will only be obtained after a sphere solution within the framework of absolute astrometry, 
e.g. the astrometric data from Gaia in its first data release (GDR1, \citealt{lindegren2016}).

The simulation is produced with AGISLab, a software package ideal for small-scale
experimental runs on a laptop that are realistic and faithful to the Gaia satellite and that 
uses a subset of the most important functionalities of the AGIS (Astrometric Global Iterative Solution) 
mainstream pipeline that is used to analyze the real Gaia data \citep{holl2012b}. 
In fact, much of the code was tested with AGISLab before being implemented in AGIS and results are 
equivalent to what can be expected with AGIS.  
For this paper the setup
is based on the actual Gaia satellite that is equipped with two field of views (FOVs) separated 
by a large `basic angle' (=106.5) rotating at a fixed spin rate of 59.9605"$s^{-1}$
around its spin axis (see \citealt{prusti2016} for extensive details). In addition, the simulator is run using nominal 
CCD size, focal plane geometry and FOV size and orbital parameters.
The observed source (proper) direction is computed using a suitable relativistic model required for high astrometric accuracies and that includes the parametrized post-Newtonian (PPN) formulation adopted for Gaia 
\citep{klioner2003} taking into account the gravitational light deflection due to solar system bodies
and the stellar aberration due to the Lorentz transformation of Gaia's co-moving reference frame 
further described in Sec.~\ref{sec:astr_lim_rel}.
 
\begin{figure*}[ht!]
\includegraphics[width=1.0\textwidth]{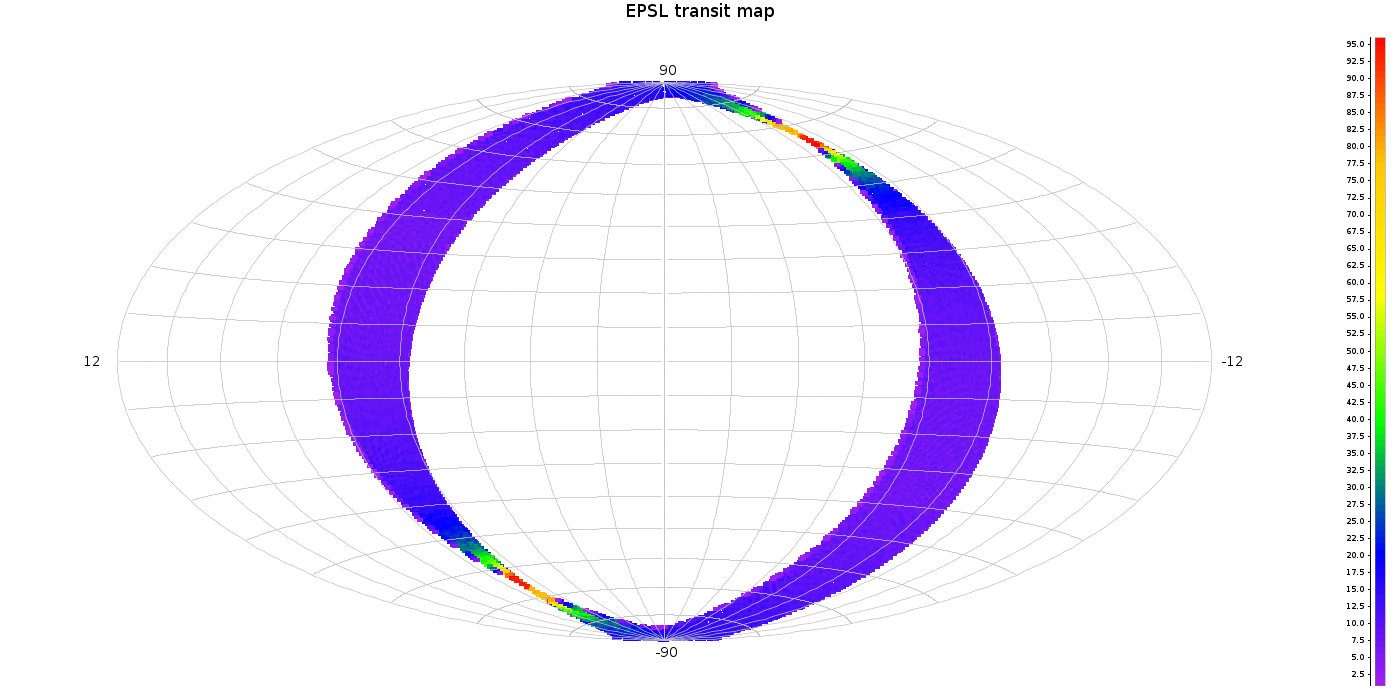}
\caption{The simulated transit map in equatorial coordinates showing the undisturbed ecliptic pole 
scanning law executed for 28 days. The different colors correspond to the number of transits quantitatively shown in
the color bar.}
\label{fig:epsl_28days}
\end{figure*}

The principle of scanning space astrometry allows 
images to be mapped onto a common focal plane thereby providing a measure of the time when 
the object transits \citep{lindbast2011}. These simulated transit times provide the 1-dimensional 
Along Scan (AL) stellar positions relative to the instrument axes.
The scanning law at a fixed solar-aspect angle is determined by two heliotropic angles, the precession 
phase, $\nu(t)$, given by the angle between the ecliptic plane and the Sun-satellite spin axis plane, and 
the spin phase, $\omega(t)$, given by the angle between the satellite's z-x plane and the 
Sun-satellite spin axis plane. The equations governing these angles has two free parameters; the 
initial spin phase and the initial precession phase at the start of the satellite science operations.

The AGISLab simulation has been setup to account for the nominal mission phase of Gaia that 
started on 25/7/2014 with the onset of an undisturbed 28 days 
Ecliptic Pole Scanning Law (EPSL) that allowed for the early calibration of several 
post-commissioning effects \citep{prusti2016}. The EPSL meant that the ecliptic poles were observed on each 
full rotation of the satellite, i.e. every 6 hours \citep{clementini2016} 
due to the precession phase being kept constant at 180$^\circ$ (see Fig.~\ref{fig:epsl_28days}).
After the EPSL the subsequent nominal scanning law was optimized in the initial spin phase 
and the initial precession 
phase to favour events of bright stars close to the limb of Jupiter for the relativity light deflection experiment due to Jupiter's quadrupole moment \citep{deBruijne2010}. 

\begin{figure*}[]
\includegraphics[width=1.0\textwidth]{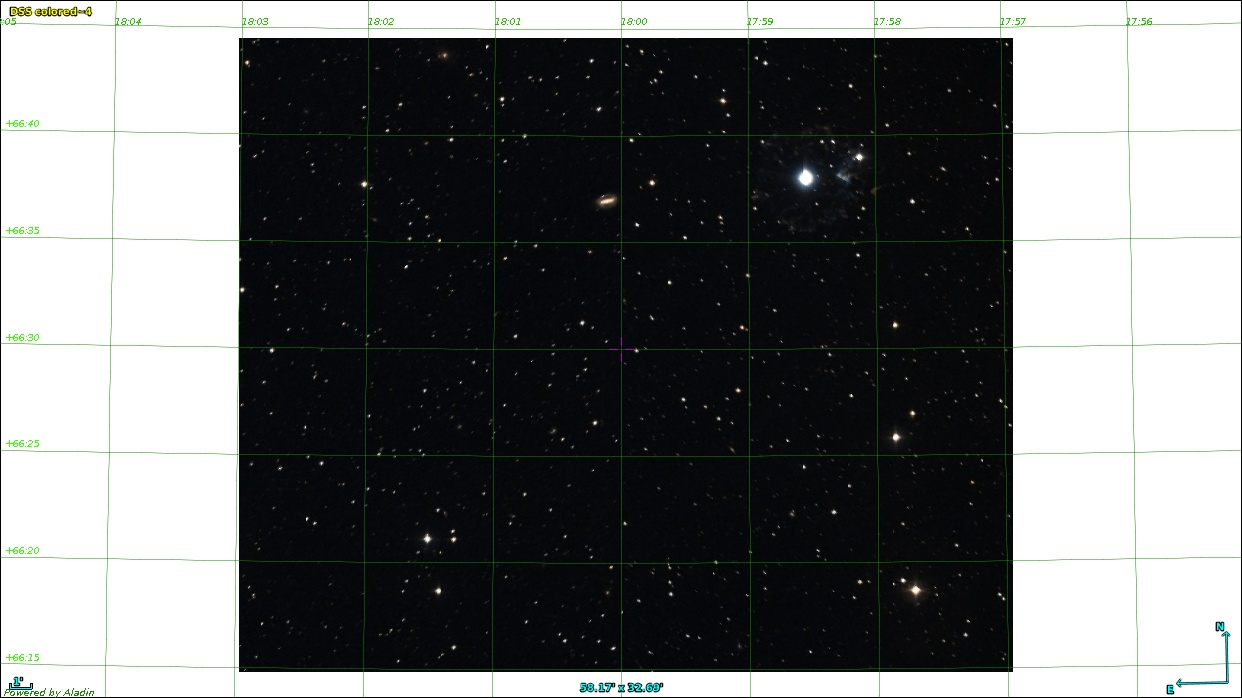}
\caption{The DSS image of the star field (size $\sim$ 0.5x0.5 degs) around the North Ecliptic Pole
courtesy Aladin.
\label{fig:dss}}
\end{figure*}

\begin{figure*}[]
\includegraphics[width=1.0\textwidth]{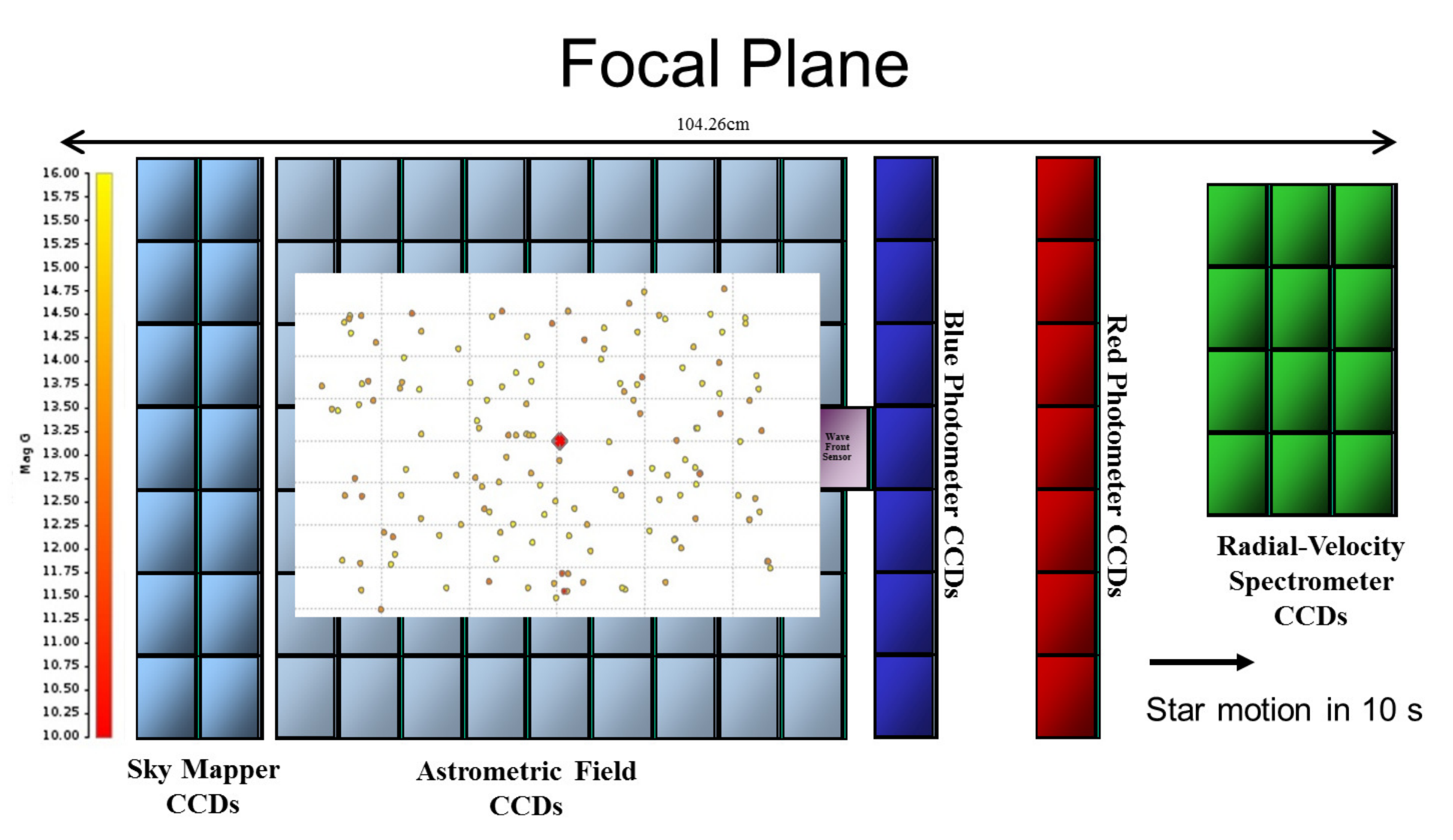}
\caption{The Gaia focal plane showing the Astrometric, Red and Blue Photometers and RV 
Spectrometer
alongwith the SkyMapper. The star field of interest is superposed on the focal plane 
and shows stars with G $<$ 16 mag in a window of size 0.6X0.4 degrees around the north ecliptic 
pole shown in red and roughly in the middle of the Astrometric Field, i.e. the fifth 
CCD column, where each grid size is 0.1 degrees in the x-direction and 0.05 degrees in the y-direction. 
The stars are colour-coded according to their magnitudes given by the vertical colour bar shown
on the left. Background image credit: ESA - A. Short. \label{fig:gaiafp}}
\end{figure*}

Taking advantage of the short time duration (which we will assume to be $\sim$ 24 hours)  
and high-cadence observations during EPSL
we can further assume that the small-scale along scan (AL) calibrations and large-scale across scan (AC)
calibrations either remain stable or are constant over long time scales, in the case of the 
former even for the whole mission duration (further described in Section~\ref{sec:instr_lim}).
Successive frames should then present a roto-translation 
between them that can be studied with the GAUSSFit software \citep{jefferys1988}, 
this is further described in the next Section.
The input stars are taken from the Initial Gaia Source List (IGSL) star catalogue 
\citep{smart2014} around the North Ecliptic Pole (NEP), this particular field is 
shown in Fig.~\ref{fig:dss}.
In order to use and fit any number of parameters we will use 
the simulated field angles in the Gaia FOV with the NEP always in the middle, 
i.e. fifth CCD column in Gaia's FOV. This is illustrated in Fig.~\ref{fig:gaiafp} where stars 
with G-magnitudes brighter than 16 in a field of size 0.6x0.4 degrees were selected and 
are shown superposed on the Astrometric Field of the Gaia Focal Plane. 
In more detail, the Focal Plane consists of a total of 106 CCDs that make up the Astrometric Field (AF), 
Blue and Red Photometer and the Radial Velocity Spectrometer. We will be concerned 
only with the Astrometric measurements provided by the 14 CCDs in the Sky Mappers (SM) and 62 CCDs in the
Astrometric Field (AF1-AF9 for a total of 9 CCD columns each with 7 rows of CCDs except for the 
last column AF9 that has 6 CCD rows).
These CCDs have 4500 and 1966 pixel columns in the AL and AC directions respectively.

In reality, Gaia will observe in Time-Delayed Integration (TDI) mode with the average 
speed of the motion of optical images (scan rate of Gaia) equal to the speed 
of charge flowing along the CCD column.
The fundamental
observational quantity is given by the time (\textit{tobs}) when a stellar image centroid  
passes the \textit{fiducial line} of a CCD, which is generally halfway between the 
first and the last TDI line used in the integration  
\citep{lindegren2012}.
As we are interested in the configuration of stars at a fixed time, the \textit{tobs} are 
converted into Along Scan (AL) positions by multiplying with the scan rate after subtraction from a reference time,  
i.e. the time of observation of the target or reference point (NEP in our case) in the same frame.

Gaia's astrometric instrument is optimized for one-dimensional measurements in the AL, whereas 
the requirements are much less stringent in the AC direction typically showing up as larger uncertainties
in AC observations versus those seen in the AL direction.
The standard uncertainty per AL/AC observation is given in 
Table~\ref{tab:ALACsig} as a function of the star's magnitude with fainter stars typically
having high uncertainties of 383 $\mu$-as in the AL-direction for G = 16 mag stars.
Generally, the AC uncertainties are 5-13 times worse for the magnitude range we will 
be looking at (G $\lesssim$ 16 mag).
For what follows it must be kept in mind that stars brighter than  G $\lesssim$ 13 mag are 
always observed as two-dimensional images that give accurate AL and AC positional information, whereas
fainter star observations acquired in the Astrometric Field are generally one-dimensional due to the AC 
position information being removed on-board by on-chip binning.
Two dimensional observations at the faint end is only sporadically available for special 
`Calibration' Faint Stars 
and instead are always provided for by the Sky Mappers (SM), 
albeit with higher uncertainties that nonetheless
provide approximate two-dimensional positions of the images.
\floattable
\begin{deluxetable}{cc}
\tablecaption{Standard uncertainties per CCD\label{tab:ALACsig}}
\tablecolumns{2}
\tablewidth{0pt}
\tablehead{
\colhead{Magnitude} &
\colhead{$\sigma$ per AL/AC observation} 
}
\startdata
G = 10 & 71 $\mu$as/367 $\mu$as\\
G = 11 & 79 $\mu$as/411 $\mu$as\\
G = 12 & 60 $\mu$as/311 $\mu$as \\
G = 13 & 95 $\mu$as/493 $\mu$as \\
G = 14 & 151 $\mu$as/1603 $\mu$as \\
G = 15 & 240 $\mu$as/2642 $\mu$as \\
G = 16 & 383 $\mu$as/4950 $\mu$as \\
\enddata
\end{deluxetable}

In order to ensure that the NEP always remains at the 
center of the FOV surrounded by the same set of stars that define the local Reference Frame, the observing 
times of the set of stars is restricted to within $\pm$15 seconds of the NEP \textit{tobs} 
for the same CCD column.
Successive observations are separated by the time it takes the star to cross from one fiducial line to the next
(approx. 4.42 secs). 
We then adopt the first configuration, i.e. \textit{tobs} of the NEP at the fiducial line of the first 
CCD column, on the first scan as the reference frame thereby obtaining the plate/CCD parameters that can 
`transform' coordinates on any other frame onto this reference. 
In order to be consistent with observations of objects that fall on the 4th row of CCDs (where the last column 
CCD is replaced by a Wave Front Sensor),
we use the first eight observing times obtained per AF transit due to the first 8 CCD columns (see Fig.~\ref{fig:gaiafp}).

\newpage
\section{Accuracy of Gaia-like space-based astrometric observations}\label{sec:limitations} 

The measurements generally are affected by relativistic systematics such as velocity aberration 
and gravitational light deflection, and by the proper motions and parallaxes of the 
sources which can be classified as physical effects. 
They are also subject to instrumental errors that would require accurate modeling of the 
instrument and its calibrations and distortions. 

\begin{figure*}[ht!]
\includegraphics[width=1.0\textwidth,trim = 18mm 60mm 18mm 60mm, clip]{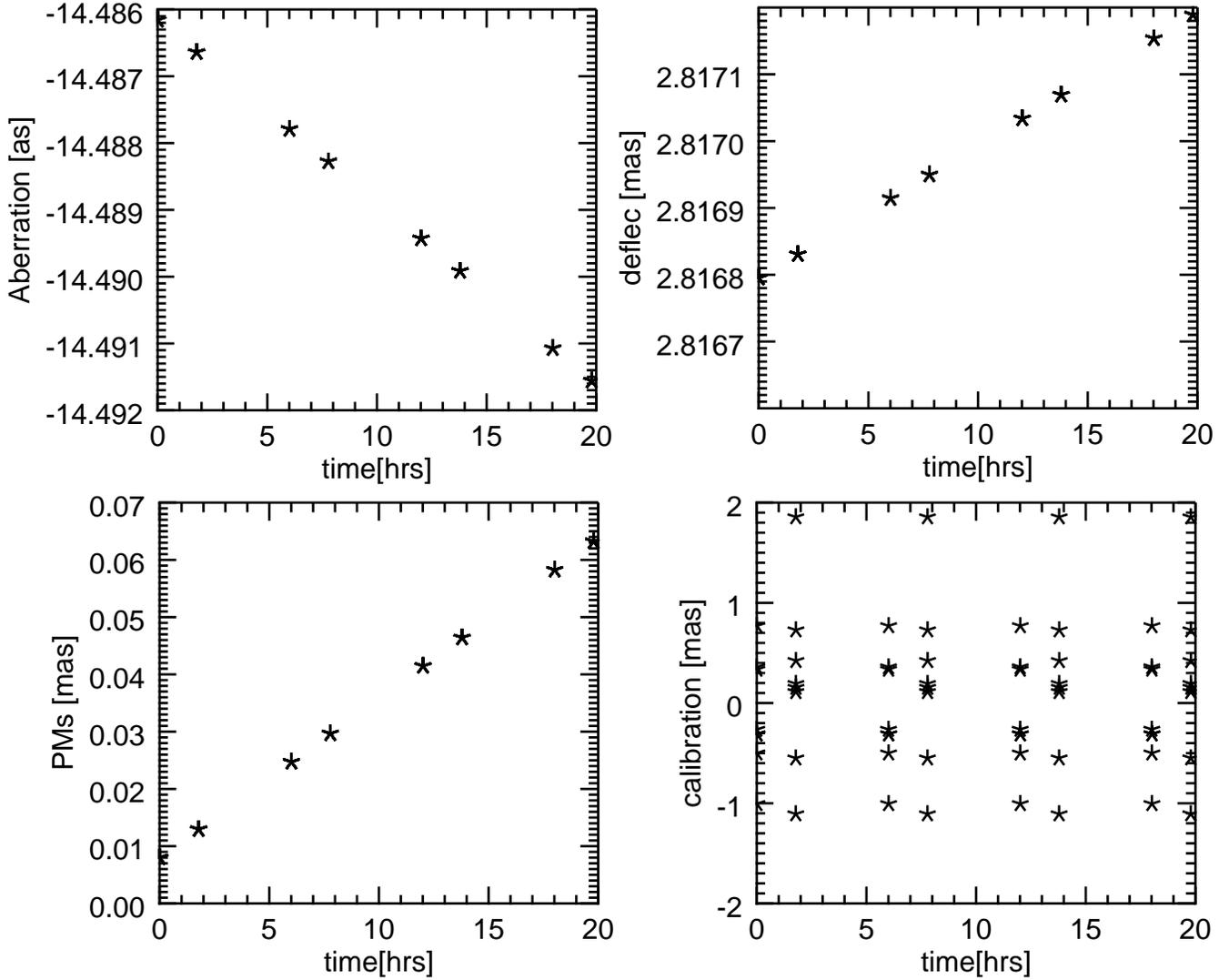}
\caption{The time dependence of the Astrometric effects averaged over all stars 
in the sample. The top panels show the time dependence of the relativistic effects 
due to the aberration (left panel) and the gravitational deflection (right panel) whereas 
the lower panels show the equivalent effects due to the star's proper motion (left panel) 
and the geometric calibrations (right panel).
}
\label{fig:diff_All}
\end{figure*}

\subsection{Physical Effects}\label{sec:astr_lim}

\subsubsection{Relativistic effects}\label{sec:astr_lim_rel}

The aberration is by far the dominant effect and is caused by the motion of the observer with 
respect to the barycenter of the solar system \citep{klioner1992, klioner2003}.
This effect can be expressed as:
\begin{align}
\delta\theta_{ab} &= \frac{v}{c} \sin\theta \left[1 + \frac{1}{c^2}(1+\gamma)w(x_0) 
		+ \frac{1}{4}\frac{v^2}{c^2} \right] \nonumber \\
	&- \frac{1}{4}\frac{v^2}{c^2} \sin 2\theta 
	 + \frac{1}{12}\frac{v^3}{c^3} \sin 3\theta  + O(c^{-4})
\end{align}
where $\theta$ is the angular distance between the direction to the target and the observer’s 
space velocity vector, $v$ is the modulus of the BCRS space coordinate velocity of the observer, 
in this case Gaia, 
$c$ is the speed of light, $w(x_0)$ is the gravitational potential of the solar system that can 
be approximated by the potential of the spherically symmetric Sun, and $\gamma$
is the parameterized post-Newtonian (PPN) parameter.
This is roughly of the order of $v/c$ to first order. For the speed of Gaia ($\simeq$ 29.6 km/s) 
the maximum values (projected values along the AL direction) are roughly 20.3" to first order, 
$\sim$2.7 mas to second order, and 3rd order terms are $\sim$ 1$\mu$as.

The gravitational deflection of light due to Solar System objects is another major effect that needs to 
be taken into account and depends on the angular separation between the Solar system body and the given 
source. This deflection due to the Sun is given by \citep{misner1973}:
\begin{eqnarray}
\delta\theta_{def} = (1 + \gamma)\frac{GM}{c^2 R_0}\cot\frac{\phi}{2} 
\end{eqnarray}
where $\phi$ is the Sun-source separation angle, $G$ is the gravitational constant, $R_0$ 
is the distance between the observer and the Sun, M is the mass of the perturbing body. 
This amounts to $\sim$ 4 mas at the NEP as seen by an earth-based observer \citep{turyshev2002}, 
and $\sim$ 2.5 mas as seen by Gaia.

Over time scales of 24 hours the  \textit{differential} aberration amounts to several mas 
whereas the \textit{differential} gravitational deflection (mainly due to the Sun's monopole) 
is sub-$\mu$as. This is shown in the top panels of Fig.~\ref{fig:diff_All} where we can see 
the strong linear 
dependence with time of the relativistic effects.

\subsubsection{Astrometric effects due to the star's proper motion}\label{sec:astr_lim_pm}

The stars' proper motions for this particular selection can vary up to several hundreds 
of mas/yr and the differential effect due to them is of the order of tens of $\mu$as 
over 24 hours as can be seen in the lower left hand panel of Fig.~\ref{fig:diff_All}. 
The IGSL input star catalog does not provide parallaxes and in this study they are  
not included in the modeling. \\

\subsection{Instrumental effects}\label{sec:instr_lim}

The observation lines, given by the fiducial lines mapped onto 
the tangent plane, are affected by the geometric instrument model describing the layout 
of the CCDs. This includes the physical geometry of each individual CCD and its configuration 
in the Focal plane assembly; the distortions and aberrations in the optical system; nominal values of
the focal length and basic angle, $\Gamma$ (see \citealt{lindegren2012, lindegren2016} 
for extensive details).
As the instrument can undergo changes during the mission, these effects are time dependent
and can be classified into three broad categories:
\begin{enumerate}
\item Large-scale AL calibrations: thermal variations in the optics, detectors and supporting structures
occuring on short time-scales and different for each FOV. 
\item Small-scale AL calibrations: physical defects or imperfections in the individual CCDs that are
expected to be stable over very long time scales, possibly over the whole mission duration. 
\item Large-scale AC calibrations: same physical origin as for 1) above, but assumed to be constant
on long time scales due to the more relaxed calibration requirement in the AC direction.
\end{enumerate}

In the differential scenario and for this paper, we will only be concerned with the Large-scale
AL and AC calibrations where the former could potentially vary over time scales of a day.
For purposes of this paper, where we use high-cadence observations over a day, we can safely assume these large scale calibrations to be constant.
The AL large-scale calibration is modeled as a low order polynomial in the across-scan pixel coordinate 
$\mu$ (that varies from 13.5 to 1979.5 across the CCD columns, \citealt{lindegren2012}) and can be written as:
\begin{equation}\label{calib}
\eta_{fn}(\mu,t) = \eta^0_n + \sum_{r=0}^2 \Delta\eta_{rfn} L_r^*(\frac{\mu-13.5}{1966})  
\end{equation}
where \textsl{f} is the field of view index, \textsl{n} is the CCD index and \textsl{r} is the degree 
of the shifted Legendre polynomial $L_r^*(\tilde{\mu})$ as a function of the normalized AC pixel coordinate 
($\tilde{\mu}$).
A similar equation holds for the AC large-scale calibrations. 
We will assume non-gated observations which is technically only valid for faint sources; brighter star
observations involve as many as a dozen gates that would need to be calibrated.  
The time-dependence of the differential calibration can be seen in Fig.~\ref{fig:diff_All} with 
a $\sim$2 mas standard deviation due to the simulated calibrations per CCD \citep{lindegren2016}. 

Furthermore, sources brighter than G $\sim$ 13 magnitude will always be observed as two-dimensional 
images, and fainter stars are observed as purely 1-dimensional images in the AF as discussed in Sec.~\ref{sec:simulation} where we will exclusively use the SM AC positional information.

\section{Applying the principles of differential astrometry} \label{sec:principles}

We adopt a differential procedure that is based on the standard one for obtaining 
the astrometric positions of the measured coordinates of an image on a plate/frame 
through the so-called plate solution. 
That usually involves using the `known positions' of the reference stars to determine
the plate solution coefficients through a least squares adjustment and then applying the plate solution 
to obtain the corresponding coordinates of the target star on the frame \citep{kovalevsky2004}. 
The inverse gnomonic projection then gives the desired astrometric coordinates
of the star. 
Here we still use the basic principle of obtaining the plate, or, for the case of Gaia, 
a frame solution, that is then used to `transform' the field angles on various FOVs to the reference FOV,
akin to `stacking' the various FOVs onto a common one.

We will use the field angles over several successive transits, for a maximum of 8 transits 
which translates into 24 hours (the crossing of a FOV represents one transit), and study the model as:
\begin{equation} \label{eqn:generic}
\begin{aligned}
x_i' &= F(\rm{frame,\ source,\ instrument\ parameters}, \mathit{x_i}) \\
     &\qquad + \mathit{offset}
\end{aligned}
\end{equation}
where $x_i'$ is the reference frame coordinate in arcseconds, which can be taken as the first 
frame field angle and $x_i$ is the `measured' field angle on the other frames. 

The overlapping frames are solved using the 
Gaussfit software \citep{jefferys1988} that is a computer program in its own 
computer language (similar to C) 
and that is especially designed for solving least squares and robust estimation problems. 
It provides a straightforward way to formulate different types of complex problems, for e.g. problems in nonlinear estimation, problems with multiple observations per equation of condition, problems with 
correlated observations etc.
It also allows the user to specify a robust estimation method that is resistant to outliers in the data.

The standard reduction method is the generalized least squares algorithm of \citep{jefferys1980}
which allows for the types of problems mentioned above alongwith their proper constraints. 
For numerical stability the solution is obtained by Householder transformation designed 
to deal efficiently with overlapping-plate conditions.

\section{The Astrometric Model} \label{sec:models}

We adopt polynomial equations for the frame model parameters to study the different effects 
outlined in Sec.~\ref{sec:limitations}.
Following Eq.~\ref{eqn:generic} from the previous section these are written as:
\begin{align} \label{eqn:poly}
x' &= a + bx + cy + dxy + ex^2 + fy^2 \nonumber \\
&+ F(\pi , \mu , \mathnormal{CCD})  \nonumber \\
y' &= g + hx + iy + jxy + kx^2 + ly^2 \nonumber \\
&+ F(\pi , \mu ,\mathnormal{CCD})
\end{align}
where $x$ and $y$ are the measured positions of the star on \textbf{any} frame, and $x'$, $y'$ are 
the positions of the stars on the \textbf{reference} frame, and $\pi$, $\mu$ are respectively 
the source parallax and proper motion.
All the measured quantities have their associated errors.
As we are looking to solve the problem in a differential manner,
we take one of the frames as the reference one, and the stars on the other frames are `adjusted'
to the reference one. The frame model equation then gives the fitted $x'$, $y'$ and consequently
how well it describes the positions of the stars. Any significant deviation is an indication that 
the model needs to be changed and adapted to the science case at hand.
With Gaussfit we can estimate the $x$ \& $y$ variables alongwith all the plate constants simultaneously.

\subsection{Linear model}
As mentioned in Sec.~\ref{sec:simulation} the observation time ranges around 
the NEP in each field of view of Gaia which then fixes the nearby distribution of 
stars subsequently simulating the field angles ($\eta$, $\zeta$) for these stars.
The linear `plate model' is given by:
\begin{align} \label{eqn:full_linPM}
\eta^{'}_{ij} &= \eta_{ij}+ \sum_{r=0}^2 \Delta\eta_{rfnk} L_r^*(\frac{\mu-13.5}{1966}) \nonumber \\
\zeta^{'}_{ij} &= \zeta_{ij}+ \sum_{r=0}^2 \Delta\zeta_{rfnk} L_r^*(\frac{\mu-13.5}{1966}) \nonumber \\
\eta^{'}_{0j} &= a_i\eta^{'}_{ij} + b_i \zeta^{'}_{ij} + c_i - \mu_{\eta j}\Delta t_{ij} - P_\eta \pi \nonumber \\
\zeta^{'}_{0j} &= d_i \eta^{'}_{ij} + e_i \zeta^{'}_{ij} + f_i - \mu_{\zeta j}\Delta t_{ij} - 
P_\zeta \pi
\end{align}
where $i$ is the frame number, and $j$ is the star number.
$\eta_{0j}$ and $\zeta_{0j}$ are the reference field angles of the $j$th star measured in the 
reference field of view, whereas $\eta_{ij}$ and $\zeta_{ij}$ are the measured nominal field angles 
of the $j$th star in the $i$th frame and their primed counterparts are the calibrated field angles. 
The constants $a_i$, $b_i$, $d_i$ and $e_i$ are scale and 
rotation plate constants, whereas $c_i$ and $f_i$ are offsets; 
$\mu_{\eta j}$, $\mu_{\zeta j}$ are the proper motions,
$\pi$ is the parallax, and $P_\eta$, $P_\zeta$ are the
computed parallax factors and $\Delta t_{ij}$ is the epoch difference between the $i$th frame star
observation and the reference frame.

For now we will also assume that the stars have zero parallaxes mainly due to the zero input values 
from the IGSL catalog.

Gaussfit solves this set of equations through a least squares procedure that minimizes the 
sum of squares of the residuals constrained by the input errors alongwith 
appropriate constraints on the proper motions \citep{eichhorn1988} 
and calibration parameters \citep{lindegren2012}. 
The fitted plate/frame parameters (a through f) then gives the 
model whereby the observations ($\eta_{ij}$, $\zeta_{ij}$) in any given frame can be `transported'
to a common reference frame. The distribution of residuals then informs us as to how well the model
accounts for various physical or instrumental effects.
It is found that $a_i$ and $e_i$ are almost unity,
whereas $b_i$ = $-d_i$  and together they give the rotation and orientation. 
The offsets $c_i$ and $f_i$ give the zero point of the common system.

\begin{figure*}[ht!]
\includegraphics[width=0.98\textwidth,trim = 5mm 20mm 20mm 90mm, clip]{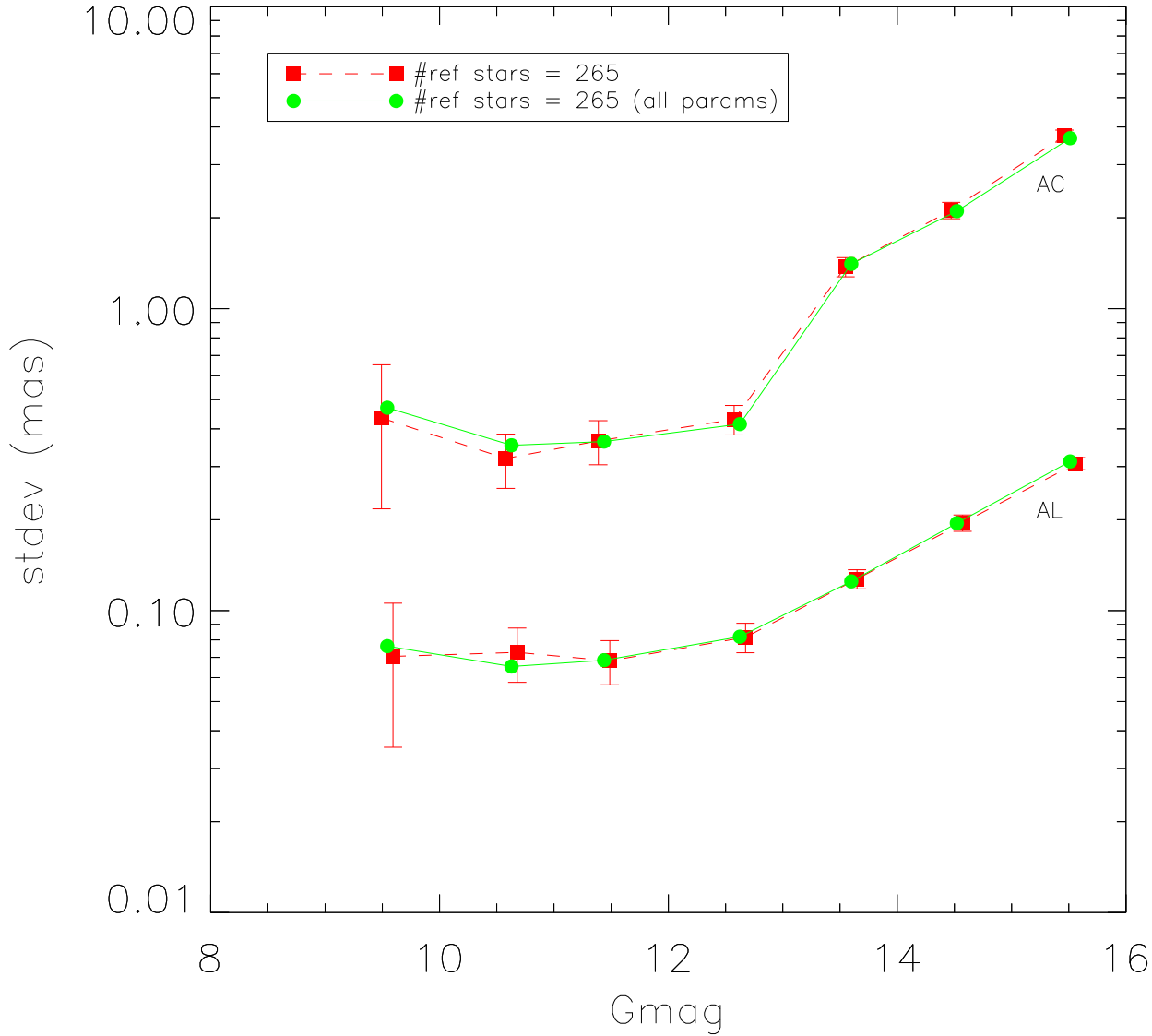}
\caption{The standard deviations (in mas) from using a linear model to fit the 
overlapping frames of observations 
around the NEP in bins of the star's G-magnitude simulated with a perfect 
instrument and no proper motions (red dashed lines) superposed to the full linear
model with all physical and instrumental effects included (green solid line). 
The lower and upper curves show the standard deviations in AL and AC respectively with input 
errors that follow the standard CCD-level location estimation errors.
The error bars are poissonian (inversely proportional to the square root of the number of stars for that
magnitude bin). }
\label{fig:stdev_GRwsig_linPM}
\end{figure*}

\subsubsection{Astrometric Reference frame residuals}\label{sec:residuals}
The standard deviation of the residuals binned as a function of the star's G-magnitude 
is shown in Fig.~\ref{fig:stdev_GRwsig_linPM}
for simulations with the Gaia Relativistic Model (GREM, \citealt{klioner2003}) 
implementation and for a perfect 
instrument and zero proper motions superposed on the full model with all effects included 
as written in Eq.~\ref{eqn:full_linPM}.
The number of reference stars is intrinsically determined by the magnitude cut 
of the sample and gives us a rough lower limit that is already according to the useful rule of
thumb where at least 3 times the number of observations is needed as coefficients to be 
determined.

\begin{figure*}[hb!]
\includegraphics[width=0.98\textwidth,trim = 8mm 10mm 20mm 80mm, clip]{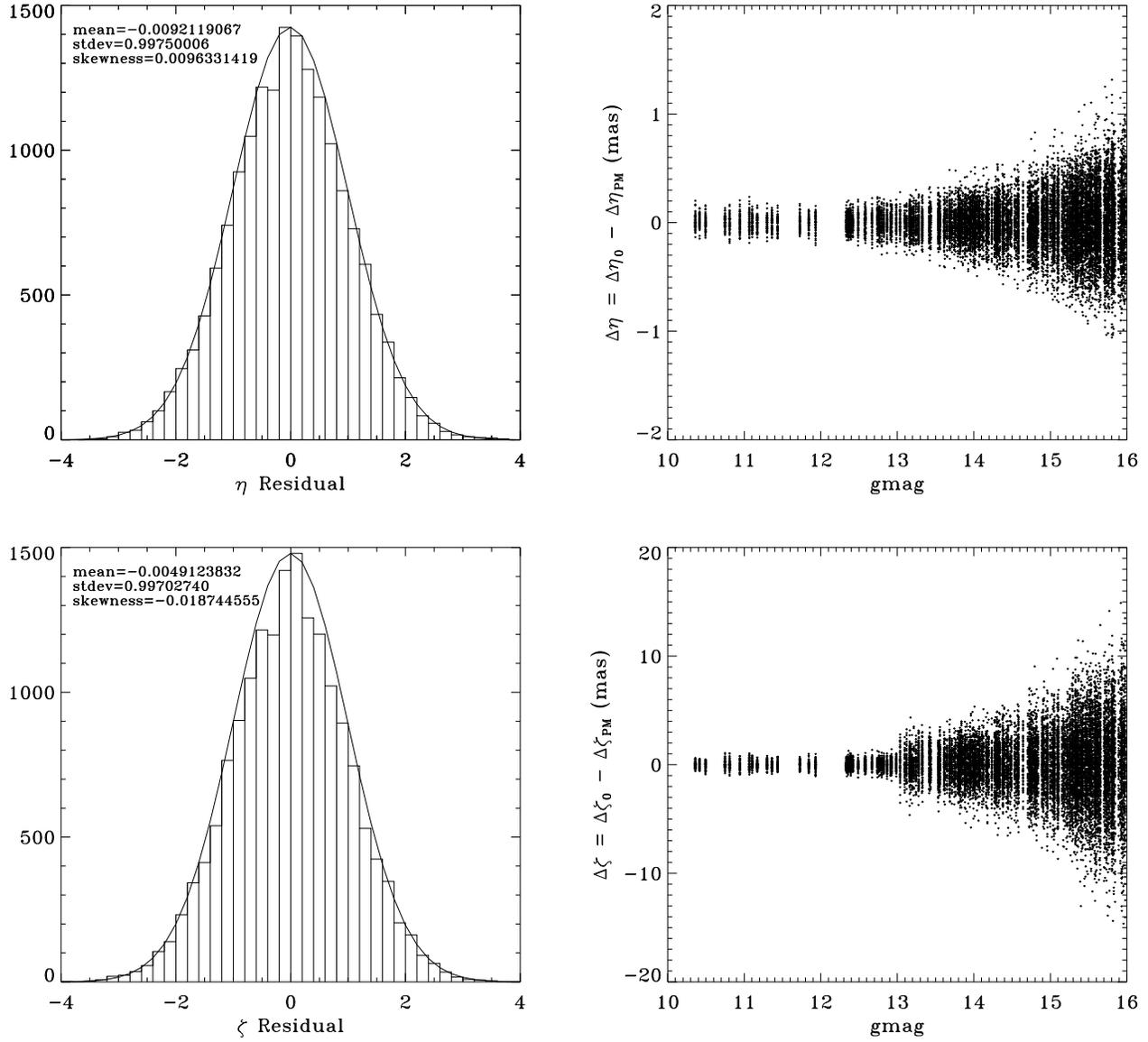}
\caption{
The histogram of the ratio of the estimated residuals and the input standard uncertainties (left panels) 
from using a linear model with proper motions and calibration parameters used to fit the overlapping 
frames of observations around the NEP for stars with G $<$ 16 mag in the AL (top panel) 
and AC directions (bottom panel). 
The fitted gaussian distributions are shown as curves overlapping the histograms.
The residuals follow a magnitude-dependent distribution (right panels) that is mainly
driven by the input errors.
}\label{fig:res_GRwPMcal_mag16}
\end{figure*}

With the addition of the geometric instrument model the number of parameters increases by a factor of
6x48x2 ([\# of calibration parameters per CCD]\textsf{x}[\# of CCDs per FOV]\textsf{x}[\# of FOVs])
$= 576$ (see Eq.~\ref{calib} and using 8 CCD columns as described in Sec.~\ref{sec:simulation} with
the added constraint of the offset between the two FOVs).
For this particular simulation we have 8 frames per transit for a total of 8 transits over 24 hours
providing us with 64 observations per star. The requirement would then be for at least 50 reference stars 
in order to reliably estimate the residuals and have a good coverage over the different CCDs.

We find a very good agreement as can be seen in Fig.~\ref{fig:stdev_GRwsig_linPM} which 
implies that a fully linear model is sufficient to describe the various physical and 
instrumental effects.

The goodness of fit can be seen in the left panels of Fig.~\ref{fig:res_GRwPMcal_mag16} 
that shows the histogram of the ratio of the residuals to the input standard uncertainties per
AL/AC observation (Table~\ref{tab:ALACsig}) for the full G $<$ 16 mag sample.
The right panels show that brighter stars have smaller astrometric residuals, as expected. We further 
tested the robustness of the astrometric solution by performing two additional tests (not shown here) that 
confirmed a direct linear dependence of the standard deviation of reconstructed test positions with 
increasing input standard uncertainties and increasing distances from the center of the reference star sample.

\newpage

\section{Discussion}\label{sec:disc}

We have studied simulated observations of Gaia field angles ($\eta$ and $\zeta$) as
measured in its FOV over short timescales of 24 hours 
within the framework of Differential Astrometry. In order to satisfy the requirement 
for a minimum number of reference stars needed to resolve the equations of 
condition we relied upon high-cadence simulated observations at the North
Ecliptic Pole such as those obtained during the EPSL (described in Sec.~\ref{sec:simulation}).
This allowed us to study the standard astrometric linear model with the inclusion of 
calibration parameters both in the AL and AC directions. 
The Gaia catalog essentially provides global astrometric products (celestial coordinates,
proper motions and parallaxes) that are obtained with the AGIS (Astrometric Global Ierative Solution) process 
involving a highly time-consuming sphere solution.
Instead, harnessing the power of differential astrometry we look at how the field angles can be used to 
construct a small field reference frame that can essentially be performed on any laptop without the need for supercomputers. 
This methodology would need to be applied to the observed field angles. 

\begin{figure}[ht!]
\includegraphics[width=0.50\textwidth,trim = 0mm 25mm 20mm 90mm, clip]{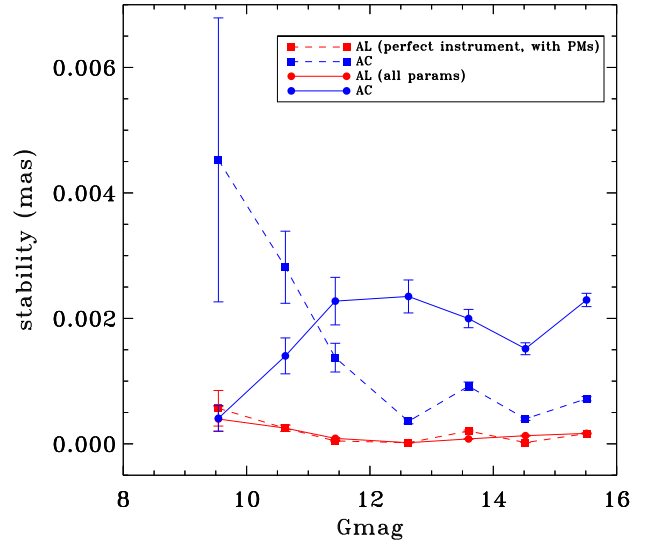}
\caption{The stability (absolute differences between the standard deviations of the 
estimated residuals and the input standard uncertainties) 
of the Differential Astrometric Reference Frame in bins of the star's G-magnitude 
for different models in the AL and AC directions with poisson errors. 
The dashed lines are for a perfect instrument model (only physical effects included), 
whereas the solid lines are for the full linear model as given in Eq.~\ref{eqn:full_linPM}. 
Red and blue lines are for the AL and AC scan directions respectively.}
\label{fig:stability}
\end{figure}

With a perfect instrument and no proper motions (only relativistic effects due to aberration 
and gravitational light deflection included) we can recover the $\mu$as stability  
of the reference frame with a modest number of stars ($\sim$ 37) down to the 13th G-magnitude 
for this particular field of interest. 
Including all astrometric effects due to physical and instrumental causes 
we require at least as many stars 
as the number of CCD's in the Gaia Focal Plane restricting us to samples with magnitude limits 
fainter than the 14th magnitude (i.e. $<$14, $<$15, $<$16). The stability is shown in 
Fig.~\ref{fig:stability} in the form of the differences between the standard deviations of the 
estimated residuals and the input standard uncertainties (Table~\ref{tab:ALACsig}) per magnitude bin. 
Moreover, the stability shows the capability of brighter stars in constraining better a target position
due to their conventionally smaller standard uncertainties.
It is remarkable that we 
are able to maintain the $\mu$as stability of the reference frame even with the inclusion of 
576 more unknowns (see Sec.~\ref{sec:residuals}).
As expected, the full linear model in AC is less stable due to the less precise positions 
in the AC direction as compared to the $\mu$as stability of the same model in AL.

Extra instrumental effects would have to be modeled as systematics and if left untouched would lead 
to a less stable small field reference frame.
The present study represents a best case scenario to show that with a slightly more simplified 
geometric instrument model we are still 
able to recover a stable small field reference frame in the AL, less so in the AC direction.
For real Gaia observations one needs to account for actual CCD gate and window effects and time dependencies 
(especially when looking at differential effects on time scales longer than just a day, 
see \citealt{lindegren2016} for more details).
Other effects due to color, magnitude, and possibly, variations in the basic angle, 
would also need to be modeled and carefully accounted for in the final processing.

\section{Future Perspectives}\label{sec:applic}

Once all linear effects due to astrometric, i.e. physical and instrumental causes, are 
accounted for in the model and a stable reference frame has been established
we can look at short and long-time effects that would 
need to be correctly modeled as a systematic effect in the differential scenario 
described. Possible experiments include:

\subsection{Relativistic Experiments}
\subsubsection{Gaia Relativistic Light Deflection Experiment on Jupiter's 
Quadrupole moment}

The nominal scanning law of Gaia after the EPSL was optimized in the 
initial parameters to favour events of bright stars
close to the limb of Jupiter for the relativity light deflection
experiment due to Jupiter's quadrupole moment (GAREQ - GAia Relativistic Experiment on 
Jupiter's Quadrupole moment, see Sec.~\ref{sec:simulation}).

Results based on simulated observations with a Galaxy model showed that Gaia 
can provide the measurement of the light bending effect due to the quadrupole moment 
with a 3$\sigma$ confidence level \citep{crosta2006}.  In that paper 
the quadrupole deflection has been parameterized by introducing a new parameter  
$\epsilon$, equal to one if  GR predictions are true. 
The total effect is formulated as a vectorial deflection angle, the sum of two contributions 
along the radial and orthoradial  directions. 
This secondary deflection has a very specific pattern as a function of 
(i) the position of the star with respect to the oblate deflector 
and (ii) the orientation of its spin axis.
A light ray grazing the limb of Jupiter would be subjected to a quadrupole induced 
deflection of $\sim$240$\mu$as superposed on a $\sim$16mas monopole term.
The aim was to assess the detection of the 
light bending effect due to the quadrupole moment of Jupiter starting from the simplest case 
(no gravitodynamical effects, no instrument model), 
providing the groundwork for future developments.  
Due to the degradation of the astrometric accuracy, this first simulation showed that 
the inclusion of stars fainter than V = 16   
does not improve the final precision on $\epsilon$. 
Instead, by running the simulation on a selection of a few epochs 
that include the maximum number of bright stars, the results showed that a single 
experiment can do almost as well as the 5-year mission. 
This gave a vital sign that more detailed investigations on specific 
bright stars spots around Jupiter were in order. 
Subsequent simulations \citep{crosta2008c, crosta2008b, crosta2008a} with selected fields 
extracted from the GSCII data base, namely a real 
count of objects around Jupiter as observed by Gaia as a function of the star's magnitude 
and distance from Jupiter's edge, singled out  
how to further improve Gaia's ability to detect the quadrupole light deflection, 
that has been predicted yet never detected.  

Due to the motion of Jupiter in its orbit such 
an event is short-lived and measurable above the $\mu$as level only for $\sim$20 hours 
for the most favourable events involving the same bright star (Gmag$<$12). 
The situation is further complicated by Gaia not necessarily 
`seeing' the bright star on successive transits.
Nonetheless, it will be interesting to see the confidence level of such effects 
on short time scales and will be the subject of future investigations. 


\subsubsection{Astrometric Gravitation Probe project}

Techniques of differential astrometry like the one described in this work and developed for the GAREQ experiment can be conveniently applied in other astrometric experiments involving astrometric tests of gravity theories. One significant example in this sense is the Astrometric Gravitation Probe (AGP) project 
\citep{2015IAUGA..2247746V,2015mas..conf..329G}, 
a concept for a space mission whose main scientific goal is the determination of the $\gamma$ and 
$\beta$ parameters of the PPN framework at the $\sim10^{-8}$ and $\sim10^{-6}$ level respectively. 
In particular, the estimation of $\beta$ would be obtained by an astrometric reconstruction of the orbit 
of Mercury, and possibily of some selected NEOs. 
Such reconstruction, requiring the determination of the ephemerides of the observed objects with respect 
to the background stars, is clearly based on techniques of differential astrometry. 
It is also expected that such objects will be observed repeatedly in partially overlapping small fields,
thus repeating the kind of situation investigated in this work. Moreover, another possible application of 
AGP is precisely a GAREQ-like observing scenario, for which a satellite 
scanning a small sky region is better suited with respect to Gaia.

\subsection{Extrasolar planets}

The promise of Gaia global astrometry in the exoplanet arena has been 
the objective of several studies in the past 15 years (e.g. \citealt{lattanzi2000, sozzetti2001,
sozzetti2014, sozzetti2016, casertano2008, perryman2014, sahlmann2015}). 
The development of a technique to extract and model very high-accuracy two-dimensional 
local astrometric measurements from the original one-dimensional Gaia data 
will allow to verify the existence of orbital motion induced by 
planetary-mass companions based on a different approach to the modeling 
of calibration and instrument attitude effects (the latter not treated 
in this work but of importance when considering time series with 
years-long time baselines). For systems for which a robust set of 
references can be established, such methodology could be particularly 
effective for the confirmation of astrometric signals corresponding to 
peculiar cases, such as edge-on, face-on, and highly eccentric orbits, 
and it might also help in the interpretation of Gaia data for very 
bright stars.  The establishment of a robust framework for the proper 
modeling of narrow-field astrometry at the $\mu$as level will also be 
valuable in the perspective of future efforts to exploit the technique 
for detection of orbital motion induced by terrestrial planets in the 
Habitable Zone of the nearest solar-type stars \citep{malbet2012}.

\subsection{Brown dwarfs}

In Gaia DR1 over 300 known ultracool later than L0  dwarfs have been 
found and from this it is estimated the final tally of L/T dwarfs directly visible to 
Gaia will be around 1000  (Smart et al, submitted). The binary fraction of L/T dwarfs 
has many published values from 10-70\% \citep[see ][ and references 
therein]{2015MNRAS.449.3651M} and the systems  visible to Gaia are all close 
so will often be resolveable. 
Very prominent examples are the nearby Epsilon Indi B A T1V+T6V binary 
\citep{2004A&A...413.1029M} and Luhman 16 a L7.5+T0.5 binary \citep{2013ApJ...767L...1L}, 
in both cases the individual objects will be resolved by, and visible to, Gaia. 
We can use the narrow field differential astrometry described here to determine 
the orbits of these systems and to
search for possible other unresolved companions such as a planet that 
has been hypothesised in the Luhman 16 system \citep{2014A&A...561L...4B}. 
The precise determination of the binary fraction remains one of the largest unknowns 
in the determination of the brown dwarf luminosity function that in turn would provide one of 
the best constraints we have on current formation theories.

\subsection{Satellite Tracking}

Another possible application is in the completely different observing scenario of the Satellite Tracking Astrometric Network (STAN). As explained in 
\cite{2015IAUGA..2236451V}, 
this project proposes the exploitation of a network of new or existing ground-based telescopes, which will be able to 
improve the orbit tracking of different types of satellites around the Earth. 
Their orbits will be determined by a continuous monitoring of their positions in sky with respect to background stars, whose positions are known with very high precision thanks to the forthcoming Gaia catalogue. 
Each satellite will be observed contemporarily from different positions from the ground, so, once again, on partially overlapping small astrometric fields.


\acknowledgments
We would like to thank the anonymous referee for several useful comments that helped 
to further improve the paper and tests that confirmed the robustness of the astrometric solution.
This work was supported by the Italian Space Agency through Gaia mission contracts - 
The Italian participation to DPAC, ASI 2014-025-R.0 and 2014-025-R.1.2015 
in collaboration with the Italian National Institute of Astrophysics.

\bibliography{astrometry}



\end{document}